\input{aipcheck}

\documentclass[
    ,draft            
  ]
  {aipproc}
\usepackage{multirow}

\layoutstyle{6x9}

\def\xte     {{\it RXTE}}

\newcommand{\src}{HETE~J1900$-$2455}
\def\apj     {{\it ApJ}}
\def\apjs     {{\it ApJS}}
\def\apjl    {{\it ApJL}}

\def\pasj    {{\it PASJ}}
\def\nat     {{\it Nature}}
\def\procspie     {{\it Proc.~SPIE}}

\newcommand{\epcs}{{\rm erg\,cm^{-2}\,s^{-1}}}
\newcommand{\eps}{{\rm erg\,s^{-1}}}
\def\la{\mathrel{\hbox{\rlap{\hbox{\lower4pt\hbox{$\sim$}}}\hbox{$<$}}}}
\def\ga{\mathrel{\hbox{\rlap{\hbox{\lower4pt\hbox{$\sim$}}}\hbox{$>$}}}}

\begin{document}

\title{Breaking the AMSP mould: the increasingly strange case of
  HETE~J1900.1$-$2455}


\author{Duncan K. Galloway}{
  address={School of Physics \& School of Mathematical Sciences, Monash
           University, VIC 3800, Australia }
}

\author{Edward H. Morgan}{
  address={Kavli Institute for Astrophysics and Space Research, MIT,
    Cambridge MA 02139, USA}
}

\author{Deepto Chakrabarty}{
  address={Kavli Institute for Astrophysics and Space Research, MIT,
    Cambridge MA 02139, USA}
}

\begin{abstract}
We present ongoing {\it Rossi X-ray Timing Explorer}\/ (\xte) monitoring
observations of the 377.3~Hz accretion-powered pulsar, \src.
Activity continues in this system more than 3~yr after discovery, at a
mean luminosity of $4.4\times10^{36}\ \eps$ (for $d=5$~kpc), although
pulsations were present only within the first 70~d. X-ray variability has
increased each year, notably with a brief interval of nondetection in
2007, during which the luminosity dropped to below $10^{-3}$ of the mean
level.
A deep search of data from the intervals of nondetection in 2005
revealed evidence for extremely weak pulsations at an amplitude of 0.29\%
rms, a factor of ten less than the largest amplitude seen early in the
outburst.

X-ray burst activity continued through 2008, with bursts typically featuring
strong radius expansion. Spectral analysis of the most intense burst detected by
\xte\/ early in the outburst revealed unusual variations in the
inferred photospheric radius, as well as significant deviations from a
blackbody. We obtained much better fits instead with a comptonisation model. 
%

%
%
\end{abstract}

\maketitle


\section{Introduction}

\src\ was
discovered following detection of a thermonuclear (type-I) X-ray burst by 
{\it HETE-II} \cite[]{vand05a}.
The 377.3~Hz pulsations were detected in a 
subsequent \xte/PCA observation \cite[]{morgan05}.
Further observations revealed
Doppler shifts of the pulse frequency,
originating from the orbital motion of the neutron star; the orbital
period is 83.25~min \cite[]{kaaret05b}.
%
%
The behaviour of \src\ since then differs remarkably from the
other AMSPs.
First, the source has been active for $>3$~yr (to July 2008), roughly an order
of magnitude longer than any other AMSP (cf. with \cite{riggio08}), and
longer even than the typical outburst intervals in those systems where it
is known (e.g. \cite{gal06b}).
Second, 
the amplitude of the pulsations
(unusually low to begin with at $<3$~\%~rms) decreased systematically on a
timescale of $\approx10$~d following several of the thermonuclear bursts
observed early in the outburst \cite{gal07a}.
Such pulse amplitude variations have not been reported in the other AMSPs
in which bursts have been detected (SAX~J1808.4$-$3658 \cite[]{chak03a}
and XTE~J1814$-$314 \cite[]{stroh03a}).  

Third, no burst oscillations have been detected 
--- in both
SAX~J1808.4$-$3658 and XTE~J1814$-$338, oscillations at the pulsar
frequency are present throughout each burst. 
%
%
Fourth,
the pulsations in \src\ were present only in the first few months
of the outburst, and have not been detected since.
While the amplitude of the pulsations in the other AMSPs may
vary throughout the outburst, they are always present when detectable,
except at the end of the outburst when the
source flux has dropped to (roughly) the background level.
For most of its active duration, \src\ has thus been essentially
indistinguishable from a faint, persistent, non-pulsing LMXB.

Here we present analyses of the ongoing \xte\/ observations of \src,
including
the long-term flux history, as well as
the results of a deep pulsation search in the 2005 data during the 
intervals in which pulsations were not detected in the individual
observations. 
We report the detection of a third burst by {\it RXTE}/PCA, as well
as two bursts by {\it INTEGRAL}\/ in 2005 and 2006, in addition to those
tabled earlier \cite{gal07a}.
Finally, we present spectral analysis of the first burst detected by \xte,
which exhibited significant devations from a blackbody spectrum.

\section{Observations}

%
Since the first six months of the outburst, during which time 
observations were more frequent, we made weekly observations of \src\ with
\xte\/ each
lasting one orbit (typical exposure $\approx3$~ks). 
We analysed data from the Proportional Counter Array (PCA), which
comprises 5 identical proportional counter units (PCU) sensitive to X-ray
photons in the range 2-100~keV, and with total effective area
$\approx6500\ {\rm cm^2}$ \cite{xte96}. Spectral resolution is
$\approx18$~\% at 6~keV; each detected photon is time-tagged to $1\mu$s.

For each observation we extracted the Standard-2 mode data for each PCU and fit
the resulting spectra separately in the 2.5--25~keV band using a
phenomenological model consisting of an absorbed blackbody plus powerlaw.
The resulting reduced-$\chi^2$ ($\chi^2_\nu=\chi^2/n_{\rm DOF}$, where
$n_{\rm DOF}$ is the number of degrees of freedom for the fit) values were
generally approximately equal to 1,
indicating a statistically acceptable fit.
We created 122-$\mu$s lightcurves covering the same energy range from
generic Event mode data, available for all but one of the observations. We
adjusted the time bins to the solary system barycenter using the JPL DE200
ephemeris 
and (for observations through
to the end of 2005) corrected for the binary orbit using the parameters of
\cite{kaaret05b}.

\section{Results}


\begin{figure}
  \includegraphics{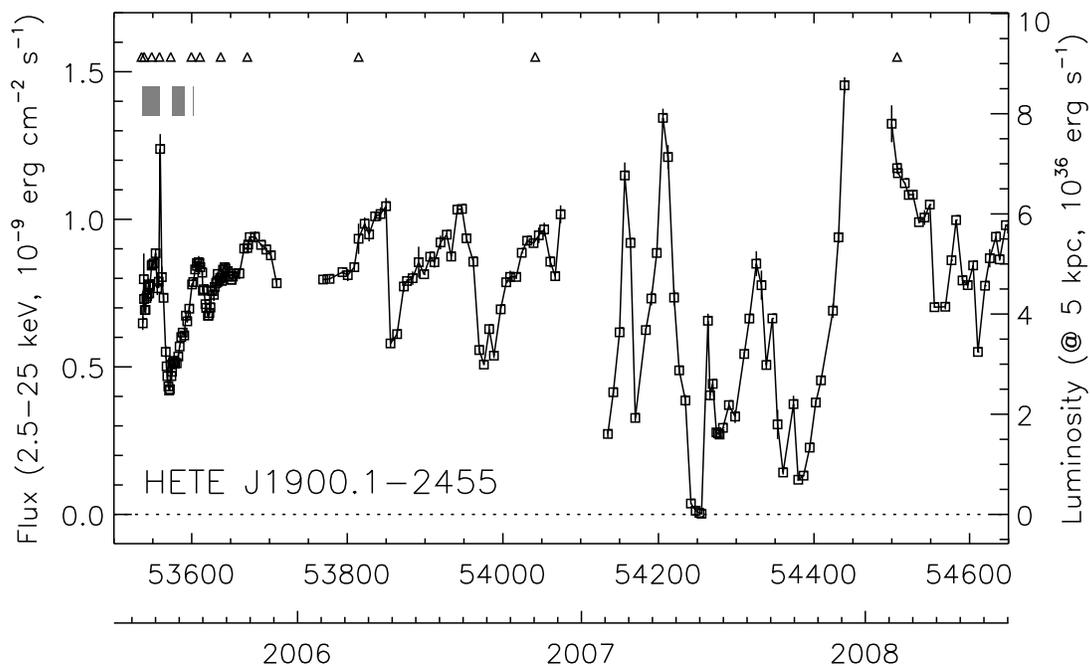}
  \caption{Long term flux history of \src, as measured by \xte. The open
squares are the fluxes measured from spectral fits of PCA observations,
averaged over the active PCUs (excluding PCU \#1). The open triangles show
the times of bursts detected by various instruments. The grey regions
indicate the intervals during which pulsations were detected.
We were unable to
observe 
for a 60-d period at the turn of each calendar year,
when the source was too close to the sun.
 \label{flux} }
\end{figure}

%

%
We measured the 2.5--25~keV flux by integrating the best-fit spectral
model over this band; the resulting flux history is shown in Figure
\ref{flux}.
The mean flux averaged over 2005--2008 June is $(7.4\pm2.5)\times10^{-10}\
\epcs$, corresponding to $4.4\times10^{36}\ \eps$ (assuming $d=5$~kpc, and
adopting a bolometric correction to the flux in the 2.5--25~keV band of
1.964), or 2.8\%~$\dot{M}_{\rm Edd}$.
This value is at least two orders of magnitude larger than the
time-averaged rate for the other AMSPs (see also \cite{gal06b}).
%
From the beginning of the outburst, we observed significant variations in
the flux on timescales of weeks--months. During 2006 these variations
appeared to repeat on a timescale of $\approx125$~d, but this was not the
case before and after. The variability of the source has increased with
each year of activity, with the possible exception of 2008. The greatest
variability in 2007 was notably accompanied by a brief transition to
quiescence; a dramatic drop in flux measured with \xte\/ in May was
followed by a non-detection by {\it Swift}\/ \cite{gal07b,deg07b}, only to
recover a few days later \cite{deg07c}. The upper limit on the luminosity
when \src\ was undetected by {\it Swift}\/ was $\sim5\times10^{32}\ \eps$;
a detailed analysis of the quiescent interval and subsequent recovery will
be presented in a subsequent paper.

%
In the monitoring observations to date we have also observed significant
variations in the X-ray colors of \src. 
These variations in some cases are associated with short-duration (of order
hours) flaring in the source; one example is the bright flare around
MJD~53559 (Fig. \ref{flux}). Such flares and related color variations have
not been observed in any of the other AMSPs; those sources exhibit a
color-color diagram typically with a single locus, without the 
large variations seen in \src.  

\subsection{Deep searches for pulsations}

%
Pulsations have not been detected in the individual \xte\/ observations of
\src\ since MJD~53602 \cite{gal07a}.
The pulsations became undetectable in two
earlier intervals, the first following a large flare early in the outburst
\cite{kaaret05b}.
The pulsations only returned
following the thermonuclear burst that was detected by the PCA on 2005
July 21 (MJD~53572). 
%
%
%
We performed a deep search for pulsations in the eight observations which
were performed between MJD~53561 and 53571 (totalling 31.3~ks). For each
observation, we performed a FFT of the orbit- and barycentre-corrected
122-$\mu$s binned, 2.5--25~keV lightcurve, and added the powers in the bin
spanning the previously determined pulse frequency \cite{kaaret05b}.
We set our detection threshold
corresponding to $3\sigma$ confidence, equivalent to a summed
(Leahy-normalised) power of 36.2. The total power detected in the bin
covering the pulse frequency was just 9.48. The corresponding upper limit
on the signal power in the summed bin is 21.3, leading to an upper limit
(at $3\sigma$ confidence) on the fractional pulse amplitude of 0.31\% rms.


Pulsations were detected in just one of five observations between
MJD~53588 and 53600. Since the total exposure during this interval was
less than in the previous nondetection interval, we cannot likely improve
on the limits determined above. Instead,
we performed a deep search for pulsations from the last detection (on
MJD~53602)
through to the end of 2005.
%
We summed FFTs from 41 observations totalling 136.7~ks 
between MJD~53608--53709. The $3\sigma$ threshold for a detection in a
single bin of the summed power spectrum is a Leahy power of 122; in the
bin corresponding to the previously measured pulse frequency of \src\ we
measured a power of 139, confirming a detection.
The
corresponding amplitude is 0.29\% rms, which is just below the upper limit
determined during the previous interval of non-detection. We also searched
each FFT individually, taking into account the number of trials, but found
no detections.



\subsection{The X-ray bursts}

Bursts from \src\ have been detected by {\it
RXTE}/PCA and ASM, {\it Swift}, and {\it INTEGRAL} (Table \ref{bursts}).
Although the majority of bursts were detected early in the outburst, this
is likely due to the increased observational duty cycle at that time; we
expect that bursting activity has continued throughout the outburst at
approximately the same level, commensurate with the relatively steady
inferred accretion rate.

The peak flux of the first burst detected by {\it HETE-II}, assuming it
reached the Eddington limit for a pure-He atmosphere, indicated a distance
of 5~kpc \cite{kawai05}; analysis of all five bursts detected by {\it
HETE-II} between 2005 June--July indicated a reduced distance of 4~kpc
\cite{suzuki07}. The first two bursts detected by {\it RXTE}/PCA had
similar peak fluxes to those detected earlier, and time-resolved spectral
analysis confirmed the identification as photospheric radius-expansion
bursts \cite{bcatalog}.
The third burst detected by {\it RXTE}/PCA on 2008 February 10 was very
similar to the second, with essentially identical peak flux, and fluence
around 10\% smaller. 
The corresponding distance (assuming that the
bursts reach the Eddington limit for pure-He material) is $4.8\pm0.5$~kpc.
As for the previous bursts detected by {\it RXTE}/PCA, we searched this
burst for oscillations at the pulse frequency, but found none.

\begin{figure}
  \includegraphics{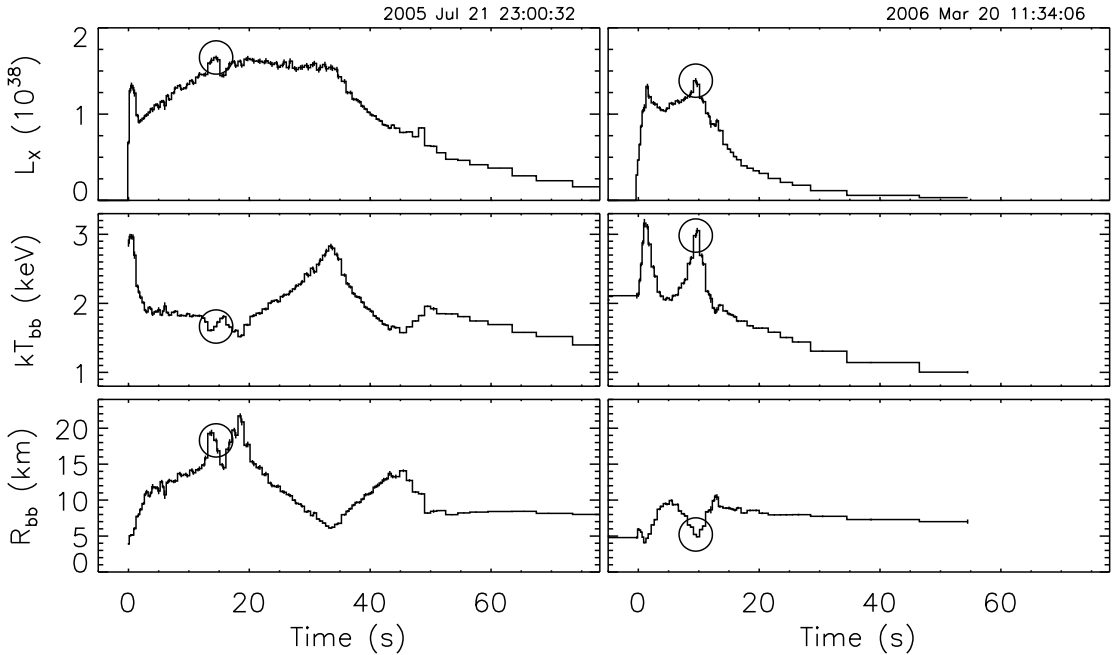}
 \caption{Two of the three X-ray bursts observed from \src\/ by \xte/PCA,
on 2005 July 21 ({\it left panels}) and 2006 March 20 ({\it right panels}). {\it
Top panels}\/ The estimated luminosity assuming isotropic emission from
a source at $d=5$~kpc. The circle marks the point at which the maximum
flux was reached. {\it Middle panels}\/ The blackbody temperature and {\it
Bottom panels}\/ The blackbody radius (at $d=5$~kpc). Error bars indicate
the $1\sigma$ uncertainties.
 \label{burst} }
\end{figure}

We also examined the spectral evolution of all three bursts detected by
{\it RXTE}. We fitted the burst spectra (using the pre-burst emission as
background,
which includes the persistent flux and the instrumental background)
with a blackbody affected by neutral absorption, following the
usual approach (e.g. \cite{bcatalog}). All three bursts exhibited radius expansion, indicated by
a local maximum in the blackbody radius near the flux maximum, coincident
with a local minimum in the blackbody temperature. The second and third
bursts, on 2006 March 20 and 2008 February 10, exhibited similar spectral
variation, both with a pronounced double peak in the (extrapolated)
bolometric flux (Figure \ref{bursts}, right panels). Although unusual,
such double peaks are not unprecedented (e.g. \cite{bcatalog}). The first
burst, however, was much more energetic, and exhibited quite bizarre
spectral evolution. A rapid rise in flux initially ended with a brief
precursor (lasting 2~s), followed by a more gradual rise to a peak at
around 15~s after the burst start (Figure \ref{bursts}, left panels).
During this rise the blackbody radius also rose steadily, but then more
sharply near the peak to give two local peaks in the radius, separated by
5~s. Following these peaks the radius decreased again, but reached a third
local maximum at around 45~s. By this time the burst flux had dropped to
approximately one-half the maximum. After this final peak the radius was
approximately constant through the remaining decay. The blackbody temperature
was roughly anticorrelated with the radius until the final decay stage.

\begin{figure}
  \includegraphics{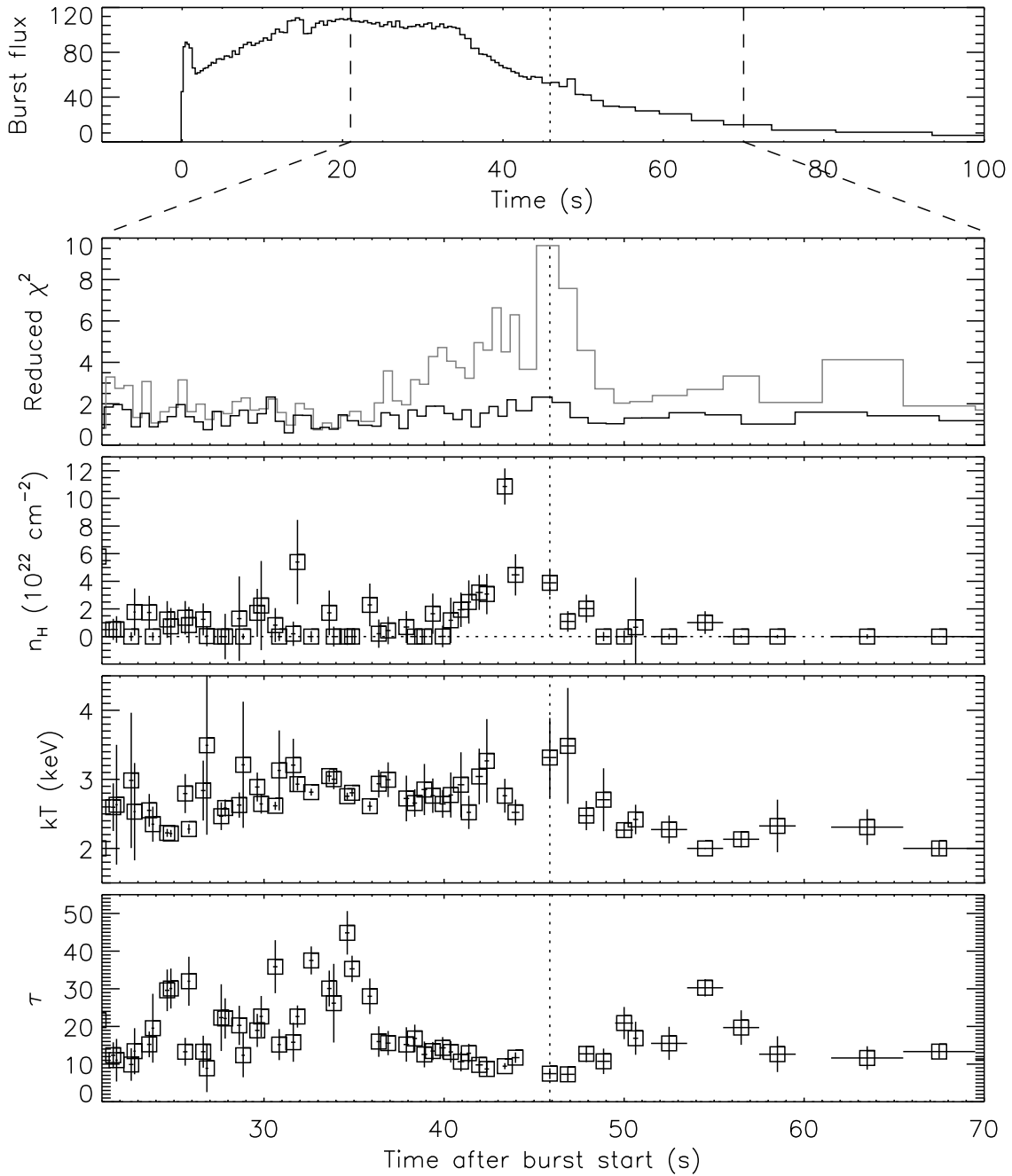}
 \caption{Spectral evolution throughout the burst from \src\ on 2005 July
21, according to an absorbed comptonisation model. The top panel shows the
burst flux (bolometric, estimated from the initial blackbody fits) in
units of $10^{-9}\ \epcs$.
The second panel compares the reduced-$\chi^2$ ($\chi^2_\nu$) values versus the blackbody
({\it grey histogram}) and comptonisation ({\it black histogram})
components during the second half of the burst peak and the initial decay.
Note the maximum $\chi^2_\nu$ of 9.6 for the blackbody fit at 45.5~s
after the burst start; this time is indicated as the vertical dotted line
in each panel.
The third panel shows the fitted neutral column density for the
comptonisation model fit; the fourth and fifth panels show the scattering
temperature $kT$ and optical depth $\tau$, respectively.
Error bars indicate the $1\sigma$ uncertainties.
 \label{comptt} }
\end{figure}

The unusual spectral evolution prompted a closer examination of the
time-resolved spectra. Throughout the first 35~s of the burst, the 
$\chi^2_\nu$ values were in the range 1--3, but during the third radius
maximum the $\chi^2_\nu$ also rose, to a maximum of greater than 9.6.
Examination of the individual spectra revealed 
a deficit of photons between 7--9~keV, suggestive of a photoionisation
feature from neutral or ionised Fe, as observed in the superburst from
4U~1820$-$30 \cite{bs04}. Using the spectral model fitted to those
data for the HETE~J1900.1$-$2455 burst, we could not obtain
an acceptable fit, likely due to the lack of strong Fe~K$\alpha$ line
emission which is expected to accompany the edge for mildly ionised disk
material. Instead, the spectra throughout the burst were well-fit with a
comptonisation model, {\tt comptt} in {\sc xspec} \cite{tit94}, with mean
$\chi^2_\nu$ values over the burst of 1.47, although with a maximum of 3.84.
The spectral fit parameters for this model show significant evolution
throughout the burst (Figure \ref{comptt}). In particular, around the time
when the $\chi^2_\nu$ for the blackbody fit reached a maximum, we find
evidence of enhanced neutral absorption, and relatively low values of
$\tau=7$--16. We are not aware of any previous analyses of burst spectra
with such a model.

\begin{table}
 \begin{tabular}{lcll} 
\hline
  \tablehead{2}{c}{b}{Start time} & 
\\
  \tablehead{1}{c}{b}{(UT)} & \tablehead{1}{c}{b}{(MJD)} &
  \tablehead{1}{c}{b}{Instrument} &
  \tablehead{1}{c}{b}{Ref.} \\
\hline
2005 Jun 14 11:22 & 53535.47361 & {\it HETE-II} 
  & \cite{vand05a,kawai05,suzuki07} \\
2005 Jun 17 21:49:10 & 53538.90914 & {\it HETE-II} 
  & \cite{suzuki07} \\ 
2005 Jun 27 13:54:10 & 53548.57928 & {\it HETE-II} 
  & \cite{suzuki07} \\ 
2005 Jul 7 13:09:22 & 53558.54891 & {\it HETE-II} & \cite{suzuki07} \\ 
\multirow{2}{*}{2005 Jul 21 23:00:32} & \multirow{2}{*}{53572.95871} &
{\it RXTE}/PCA \& & \cite{gal07a} \\ 
  & & {\it HETE-II} 
  & \cite{suzuki07} \\
2005 Aug 17 12:19:58 & 53599.51387 & {\it Swift}\tablenote{ BAT triggers
\#150823, 152451; C. Markwardt,
pers. comm. (2005)} 
  & \\ 
2005 Aug 28 15:09:37 & 53610.63167 & {\it Swift}$^*$ 
  & \\ 
2005 Sep 24 04:47:10 & 53637.19941 & {\it RXTE}/ASM\tablenote{R.
Remillard, pers. comm. (2005)} \\ 
2005 Oct 28 10:25:30 & 53671.43438 & {\it INTEGRAL}\tablenote{Events
  2693/0, 3529/0 at 
http://wydra.ncac.torun.pl/$\sim$jubork/ibas } \\
2006 Mar 20 11:34:06 & 53814.48202 & {\it RXTE}/PCA 
  & \cite{gal07a} \\ 
2006 Nov 2 13:32:22 & 54041.56414 & {\it INTEGRAL}$^{**}$ \\
2008 Feb 10 20:32:51 & 54506.85615 & {\it RXTE}/PCA \\
\hline
\end{tabular}
\caption{Type-I X-ray bursts observed from \src
  \label{bursts} }
\end{table}


\section{Summary \& future prospects}

\src\ continues to reveal surprising behaviour. The
``failed'' transition to quiescence in 2007, as well as the X-ray
intensity and color variability, further extends the range of AMSP
phenomenology. We have also found unusual burst spectral shape and
variations in the first burst detected by \xte. Significant deviations
from the usual blackbody model could not be modelled using a reflection
spectrum, but instead were fit with an absorbed comptonisation model.
During the stage of the burst in which we detected the most significant
devations from a blackbody spectrum, we found evidence of significantly
enhanced neutral absorption, as well as a lower-than-usual optical depth
for scattering $\tau$.
It is not clear how (if at all) the unusual spectral shape and variations
may be related to the atypical behaviour inferred for the photospheric
radius throughout the expansion.  A preliminary study of spectra from
almost 1200 bursts observed by \xte\/ \cite{bcatalog} indicates that
deviations from a blackbody model can be most commonly explained by
comptonisation, both in radius-expansion and non-radius expansion bursts. 

However, it is the behaviour of the pulsations early in the outburst that
is of most interest. 
The accumulated exposure from the last detection
through to the end of 2005 allowed us to detect a weak signal with amplitude
of just 0.29\%~rms. At just below the upper limit obtained for the first
interval of nondetection, we cannot exclude the presence of pulsations at
this level also in that interval.  Rather than being absent altogether,
the pulsations appear to be present but at an amplitude of around 10\% of
the maximum reached earlier in the outburst. 
For the observations from 2006 onwards, the accumulated error in the
orbital ephemeris necessitates an acceleration search for pulsations. 
This search is currently under way.

The lack of pulsations in the broader class of LMXBs
(which are mostly accreting above the rates typical for the AMSPs)
has been suggested to arise from ``burial'' of the magnetic field by the
accreted material. 
A combination of properties in \src\ (unusually low
magnetic field strength, sustained accretion) perhaps makes it uniquely
susceptible to burial compared to the other AMSP systems
(see also Cumming, this volume).
The continued presence of pulsations at $\approx10$\% of the peak
amplitude detected earlier in 2005 suggests that the field may not be
completely buried, but reduced by a commensurate amount.

If such mechanisms prove plausible to explain the properties of the
pulsations in \src, possibly including the response to the thermonuclear
bursts, the future behaviour --- particularly episodes of very low
accretion rate, as was observed in 2007 --- may provide direct tests of
the burial scenario.



\begin{theacknowledgments}
We thank the workshop organisers for an interesting and enjoyable
conference program.
\end{theacknowledgments}




\begin{thebibliography}{19}
\expandafter\ifx\csname natexlab\endcsname\relax\def\natexlab#1{#1}\fi
\providecommand{\enquote}[1]{``#1''}
\expandafter\ifx\csname url\endcsname\relax
  \def\url#1{\texttt{#1}}\fi
\expandafter\ifx\csname urlprefix\endcsname\relax\def\urlprefix{URL }\fi
\providecommand{\eprint}[2][]{\url{#2}}

\bibitem[{Vanderspek} et~al.(2005)]{vand05a}
R.~{Vanderspek}, E.~{Morgan}, G.~{Crew}, C.~{Graziani}, and M.~{Suzuki},
  \emph{The Astronomer's Telegram} \textbf{516} (2005).

\bibitem[{Morgan} et~al.(2005)]{morgan05}
E.~{Morgan}, P.~{Kaaret}, and R.~{Vanderspek}, \emph{The Astronomer's Telegram}
  \textbf{523} (2005).

\bibitem[{Kaaret} et~al.(2006)]{kaaret05b}
P.~{Kaaret}, E.~H. {Morgan}, R.~{Vanderspek}, and J.~A. {Tomsick}, \emph{\apj}
  \textbf{638}, 963--967 (2006).

\bibitem[{Alpar} et~al.(1982)]{alpar82}
M.~A. {Alpar}, A.~F. {Cheng}, M.~A. {Ruderman}, and J.~{Shaham}, \emph{\nat}
  \textbf{300}, 728--730 (1982).

\bibitem[{Radhakrishnan} and {Srinivasan}(1982)]{rs82}
V.~{Radhakrishnan}, and G.~{Srinivasan}, \emph{Current Science} \textbf{51},
  1096--1099 (1982).

\bibitem[{Riggio} et~al.(2008)]{riggio08}
A.~{Riggio}, T.~{Di Salvo}, L.~{Burderi}, M.~T. {Menna}, A.~{Papitto},
  R.~{Iaria}, and G.~{Lavagetto}, \emph{\apj} \textbf{678}, 1273--1278 (2008).

\bibitem[{Galloway}(2006)]{gal06b}
D.~K. {Galloway}, \enquote{Accretion-powered Millisecond Pulsar Outbursts,} in
  \emph{The Transient Milky Way: a perspective for MIRAX}, edited by
  F.~{D'Amico}, J.~{Braga}, and R.~{Rothschild}, AIP, Melville, NY, 2006.

\bibitem[{Galloway} et~al.(2007{\natexlab{a}})]{gal07a}
D.~K. {Galloway}, E.~H. {Morgan}, M.~I. {Krauss}, P.~{Kaaret}, and
  D.~{Chakrabarty}, \emph{\apjl} \textbf{654}, L73--L76
(2007{\natexlab{a}}).

\bibitem[{Chakrabarty} et~al.(2003)]{chak03a}
D.~{Chakrabarty}, E.~H. {Morgan}, M.~P. {Muno}, D.~K. {Galloway},
  R.~{Wijnands}, M.~{van der Klis}, and C.~B. {Markwardt}, \emph{\nat}
  \textbf{424}, 42--44 (2003).

\bibitem[{Strohmayer} et~al.(2003)]{stroh03a}
T.~E. {Strohmayer}, C.~B. {Markwardt}, J.~H. {Swank}, and J.~{in 't Zand},
  \emph{\apjl} \textbf{596}, L67--L70 (2003).

\bibitem[{Jahoda} et~al.(1996)]{xte96}
K.~{Jahoda}, J.~H. {Swank}, A.~B. {Giles}, M.~J. {Stark}, T.~{Strohmayer},
  W.~{Zhang}, and E.~H. {Morgan}, \emph{\procspie} \textbf{2808}, 59--70
  (1996),

\bibitem[{Galloway} et~al.(2007{\natexlab{b}})]{gal07b}
D.~{Galloway}, E.~{Morgan}, D.~{Chakrabarty}, and P.~{Kaaret}, \emph{The
  Astronomer's Telegram} \textbf{1086} (2007{\natexlab{b}}).

\bibitem[{Degenaar} et~al.(2007{\natexlab{a}})]{deg07b}
N.~{Degenaar} {et~al.}
  \emph{The Astronomer's Telegram} \textbf{1098} (2007{\natexlab{a}}).

\bibitem[{Degenaar} et~al.(2007{\natexlab{b}})]{deg07c}
N.~{Degenaar} {et~al.}
  \emph{The Astronomer's Telegram} \textbf{1106}
  (2007{\natexlab{b}}).

\bibitem[{Kawai} et~al.(2005)]{kawai05}
N.~{Kawai}, M.~{Suzuki}, {for the HETE Team}, \emph{The Astronomer's
  Telegram} \textbf{534} (2005).

\bibitem[{Suzuki} et~al.(2007)]{suzuki07}
M.~{Suzuki}, N.~{Kawai}, {for the HETE Team},
  \emph{\pasj}
  \textbf{59}, 263--268 (2007).

\bibitem[{Galloway} et~al.(2008)]{bcatalog}
D.~K. {Galloway}, M.~P. {Muno}, J.~M. {Hartman}, P.~{Savov}, D.~{Psaltis}, and
  D.~{Chakrabarty}, \emph{\apjs, accepted (astro-ph/0608259)}  (2008).

\bibitem[{Ballantyne} and {Strohmayer}(2004)]{bs04}
D.~R. {Ballantyne}, and T.~E. {Strohmayer}, \emph{\apjl} \textbf{602},
  L105--L108 (2004).

\bibitem[{Titarchuk}(1994)]{tit94}
L.~{Titarchuk}, \emph{\apj} \textbf{434}, 570--586 (1994).

\end{thebibliography}

\IfFileExists{\jobname.bbl}{}
 {\typeout{}
  \typeout{******************************************}
  \typeout{** Please run "bibtex \jobname" to optain}
  \typeout{** the bibliography and then re-run LaTeX}
  \typeout{** twice to fix the references!}
  \typeout{******************************************}
  \typeout{}
 }

\end{document}